# Simulating device-to-device communications in OMNeT++ with SimuLTE: scenarios and configurations

Giovanni Nardini, Antonio Virdis, Giovanni Stea
Dipartimento di Ingegneria dell'Informazione, University of Pisa
Largo Lucio Lazzarino 1, I-56122, Pisa, Italy
g.nardini@ing.unipi.it, a.virdis@iet.unipi.it, giovanni.stea@unipi.it

*Abstract*—SimuLTE is a tool that enables system-level simulations of LTE/LTE-Advanced networks within OMNeT++. It is designed such that it can be plugged within network elements as an additional Network Interface Card (NIC) to those already provided by the INET framework (e.g. Wi-Fi). Recently, device-to-device (D2D) technology has been widely studied by the research community, as a mechanism to allow direct communications between devices of a LTE cellular network. In this work, we present how SimuLTE can be employed to simulate both one-to-one and one-to-many D2D communications, so that the latter can be exploited as a new communication opportunity in several research fields, like vehicular networks, IoT and machine-to-machine (M2M) applications.

## I. INTRODUCTION

In recent years, OMNeT++ [1] has become the reference framework for performing simulation, especially in the networking field. Several OMNeT++-based models are available to support researchers in assessing the performance of networking systems, e.g. INET [2]. SimuLTE is one of those models, which provides a simulation framework for the data plane of LTE-Advanced (LTE-A) networks [3]. A peculiarity of SimuLTE is its Network Interface Card (NIC) implementation: similarly to other models available in the INET framework (e.g. Wi-Fi) it can be easily integrated within any other OMNeT++ module, so as to endow nodes with LTE connectivity. It also embeds the possibility to simulate device-to-device (D2D) communications. D2D is a new communication paradigm that enables proximity services for cellular users, since it allows two or more terminals to communicate directly, without using the base station as a relay. In particular, both one-to-one and one-to-many D2D communications can support new services for a variety of scenarios, such as vehicular networks, Internet of Things (IoT) or machine-to-machine (M2M) applications [7]. For this reason, it has attracted increasing attention from the research community. Link-level simulations are typically used for testing channel properties of D2D links, also due to the substantially lack of tools for evaluating system-level performance of D2D. In this regard, SimuLTE enhanced with D2D may play a crucial role in allowing researchers to assess the performance of new algorithms and services in several scenarios and configurations.

In this paper, we describe how to setup simulations with D2D in SimuLTE, with a special focus on the main parameters that might affect the performance of the system. We also discuss how different settings of those parameters can be employed to evaluate different scenarios.

The remainder of the paper is organized as follows. An overview on D2D technology in LTE-Advanced networks is given in Section II. Section III presents the architecture of SimuLTE, enhanced to support D2D functionalities. Section IV describes the main parameters and the possible configurations for simulating one-to-one and one-to-many D2D communications. Section V concludes the paper.

## II. DEVICE-TO-DEVICE COMMUNICATIONS IN LTE

With reference to Figure 1, in a conventional LTE-Advanced network, every communication from/to cellular devices, called User Equipments (UEs), occurs only with the serving base station, which is called eNodeB (eNB), in either the downlink (DL) or the uplink (UL) direction. Recently, device-to-device (D2D) communications have been introduced in the LTE-Advanced standard [4]. D2D allows proximate UEs to exchange data directly using the so-called *sidelink* (SL) path, avoiding the two-hop path through the eNB, hence reducing latency and saving resources. Both *one-to-one* (or unicast) and *one-to-many* (or multicast) D2D communications are envisaged. In the one-to-one case, there is a peering session established between two UEs, which communicate in the same way as they would with the eNB. In the one-to-many case, the sender UE transmits data using a MAC-level identifier for the group of UEs that should receive the message. However, in both cases, the eNB still controls the allocation of resources, in either a (semi-)static or a dynamic way (e.g. assigning resource grants to UEs upon reception of a scheduling request). Hybrid Automatic Repeat reQuest (H-ARQ) feedback and retransmissions are supported for the unicast case only.

Both unicast and multicast D2D communications present peculiar challenges that are currently being studied by the research community, such as resource allocation. System-level simulations aid researchers to investigate potential solutions to those issues, hence we provide a framework that enables the simulation of D2D-capable devices in different scenarios and under different settings of the system parameters.

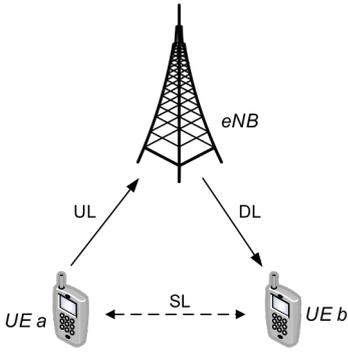
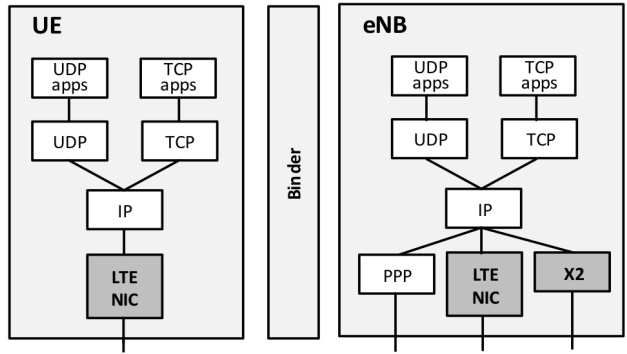
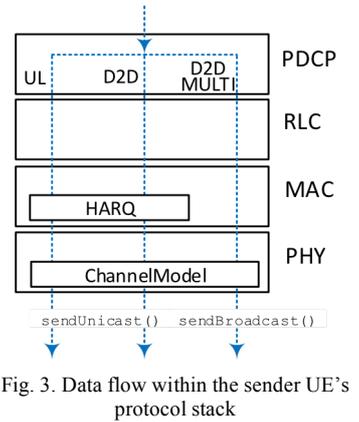

Fig. 1. Device-to-device communications

Fig. 2. SimuLTE architecture

Fig. 3. Data flow within the sender UE's protocol stack

## III. THE SIMULTE FRAMEWORK

This section provides a background on the architecture of SimuLTE, a system-level simulator based on the OMNeT++ and INET frameworks. The basic concept of OMNeT++ is *modularity*: simulations are built by composing different modules that interact with each other using messages. INET provides a large amount of entities and protocols for simulating both wired and wireless networks. In particular, it provides the concept of Network Interface Card (NIC) modules. The latter can be integrated within other modules so as to model different communication protocols between network devices. For example, devices can be endowed with Wi-Fi or Point-to-Point Protocol (PPP) NIC modules, or both. SimuLTE exploits this feature, as it is designed as an extension of a wireless NIC module. This allows one to add LTE capabilities to a node included in the simulation.

SimuLTE provides models for both UEs and eNB. As shown in Figure 2, both nodes contain the LTE NIC, together with modules implementing upper layer protocols, taken from INET. In addition, the eNB has an interface to the Internet via PPP and can also be connected to other eNBs using the X2 interface. The LTE NIC in both the UE and the eNB implements the whole LTE protocol stack, as one submodule per layer, namely Packet Data Convergence Protocol (PDCP), Radio Link Control (RLC), MAC and PHY. Since UEs and eNBs perform different operations within the protocol stack, SimuLTE exploits the inheritance paradigm of OMNeT++ for defining both the structure and behavior of each submodule. In particular, each submodule has common operations, which are extended with functionalities specific for the UE and the eNB, respectively. Communication between different layers occurs via message exchange, same as data transmission between UEs and eNBs. On the other hand, resource accounting is decoupled from data transmission. A dedicated module, called *Binder*, monitors which resources, i.e. Resource Blocks (RBs), are used by both the eNBs (for downlink transmissions) and the UEs (for uplink transmissions). The Binder can be considered as the oracle of the LTE network, since all the LTE nodes can access it to share common information via direct method calls.

Air transmissions between LTE NICs are modeled by the *ChannelModel* class, included within the PHY layer of the LTE NIC itself. On reception of a new message, the ChannelModel computes the Signal-to-Interference-and-Noise Ratio (SINR) perceived by the node. To do this, it obtains information from the Binder about the usage of the RBs for all the nodes in the network and decides whether the message can be successfully decoded or not. ChannelModel is also responsible for computing and reporting the Channel Quality Indicator (CQI) of the UEs, which is used for scheduling operations at the eNB. SimuLTE comes with a realistic implementation of the ChannelModel that takes into account path loss and fading effects. However, it can be easily re-implemented to allow one to employ its own preferred model.

We enhanced SimuLTE so as to endow the LTE NIC with D2D capabilities, enabling both one-to-one [5] and one-to-many direct communications between UEs. From the UE's perspective, most D2D operations are common to those performed for traditional UL transmissions, thus we implemented a new module per LTE layer, which inherits common functionalities from UL ones and adds D2D-specific functionalities, e.g. reporting of CQI on the SL path. With reference to Figure 3, data flows are forked at the PDCP level when they arrive from upper layers, based on whether the transmission direction is UL, D2D or D2D_MULTI. In the latter case, each packet contains also an identifier for the multicast group that must receive the packet. At the lower layers, each flow is subjected to different elaborations according to its direction. For example, in the multicast case, packets cannot use H-ARQ functionalities. When the packet reaches the PHY layer, a message is sent to the receiving UE using a `sendDirect()` call. As for the multicast case, the `sendBroadcast()` function sends a copy of the message to all UEs within the transmission range of the sender, exploiting the implementation of wireless broadcast transmissions of INET. At the reception side, UEs discard the message if they are not subscribed to the relevant multicast group.

## IV. SIMULATING D2D WITH SIMULTE

SimuLTE natively provides models of UEs and eNBs, hence it is possible to simulate D2D communications within cellular networks, possibly mixed with other Internet devices, e.g. taken from INET. However, the NIC implementation of SimuLTE described in the previous section can be fit to any network device. For example, *VeinsLTE* [6] uses SimuLTE to simulate heterogeneous vehicular networks, where vehicles are equipped with both Wi-Fi and LTE NICs. Following this direction, it is possible to design nodes with multiple network interfaces, e.g. LTE, Wi-Fi, Bluetooth etc., and compare D2D with other technologies for short-range communications.

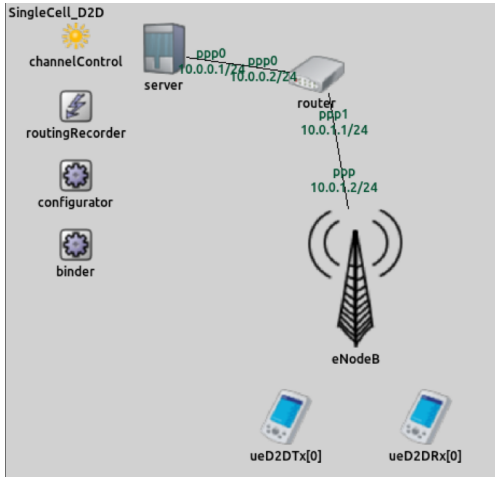

Fig. 4. Example scenario for one-to-one D2D communications

We now describe how to configure D2D-related simulation parameters in the .ini configuration file and how different parameters can be employed for evaluating different scenarios.

*A. One-to-one D2D*

Figure 4 reports an example scenario for the unicast case, where ueD2DTx[0] and ueD2DRx[0] are the sender and the receiver of the flow, respectively. A snippet of the omnetpp.ini file is reported below.

```
# enable D2D capabilities
*.eNodeB.d2dCapable = true
*.ueD2D*[*].d2dCapable = true

# select the AMC mode
*.eNodeB.nic.mac.amcMode = "D2D"

# set peering relationship
*.ueD2DTx[0].nic.d2dPeerAddresses="ueD2DRx[0]"

# select the CQI for D2D transmissions
*.eNodeB.nic.phy.enableD2DCqiReporting = false
**.usePreconfiguredTxParams = true
**.d2dCqi = 7

# set Tx Power
*.ueD2DTx[0].nic.phy.ueTxPower = 26   # in dB
*.ueD2DTx[0].nic.phy.d2dTxPower = 20  # in dB
```

In order to enable D2D communication between the two UEs, they must be defined as D2D-capable UEs, as well as the eNB. To do this, the d2dCapable parameter of the involved nodes has to be set in the .ini file. For the eNB, it is necessary to use the D2D-specific Adaptive Modulation and Coding (AMC) module, for computing transmission parameters for the SL. Then, we need to configure the receiver as a possible D2D peer for the sender. To this aim, the d2dPeerAddresses parameter is a blank spaces-separated list of IP addresses or aliases of UEs. Note that peering is unidirectional, hence we need to explicitly define the reverse one for bidirectional flows, e.g. TCP connections.

As far as CQI computation is concerned, it is possible to use fixed CQIs for every D2D transmission, set at the beginning of the simulation. Otherwise, it is possible to let sender UEs report the CQI measured on the D2D links, for all their peering UEs, i.e. those in the d2dPeerAddresses list. The latter mode allows one to obtain link adaptation of the SL, as for UL and DL communications. On the other hand, this method is computationally heavier, since CQI reporting may be done periodically and the number of peering UEs may be high. In the example we use fixed CQI, thus we set the usePreconfiguredTxParams parameter and define the CQI to be used, i.e. **.d2dCqi=7. In this case, the enableD2DCqiReporting parameter is set to false. If both modes are enabled, preconfigured CQI would be used, and reporting would be useless and would only waste computations. At the PHY layer, it is possible to employ different transmission power for UL and D2D. To do this, ueTxPower and d2dTxPower parameters should be set accordingly. This allows one to evaluate, for example, the maximum distance or the consumed energy for a direct communication under specific combinations of CQI and transmission power. Since D2D enables spatial frequency reuse, different settings also affects interference among UEs allocated in the same resources, hence it is possible to assess the performance of scheduling algorithms that exploit frequency reuse and the coexistence between D2D and traditional infrastructure communications.

The communication mode between two D2D-capable UEs may be changed during the simulation, from D2D mode to the traditional infrastructure mode. In fact, UEs may get too far from each other, thus reducing the SINR on the SL, or the eNB may want to optimize the allocation of resources according to its own policy. This topic has recently gained attention in the research community, e.g. [8][9], and the need to simulate those algorithms may arise. To this aim, we endow the LTE NIC at the eNB with a module called d2dModeSelection. It is implemented as an abstract class that periodically invokes the doModeSelection() function. The latter runs an algorithm that, for each D2D-capable UEs, selects which mode has to be be used in the next period to communicate with each of its possible receivers, i.e. those in the d2dPeerAddresses list. Since doModeSelection() is a pure virtual function, it must be re-implemented by derived classes, as shown in Figure 5. The d2dModeSelectionBestCqi module implements a simple algorithm that chooses either D2D or infrastructure mode based on the best CQI between the UL and the SL. This architecture allows one to realize its own mode selection policy by extending the base module and implementing the

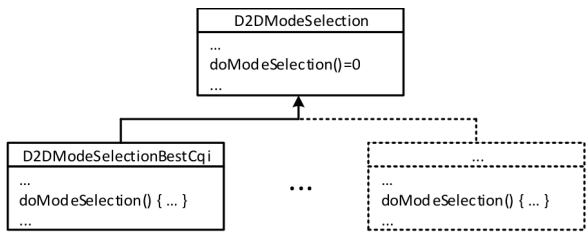

Fig. 5. D2D Mode Selection architecture

`doModeSelection()` function. When the selection has been made for each possible D2D pair, a control message is sent to both the endpoints to command the switch to the new communication mode. This operation is similar to the handover and ongoing communications between the UEs involved in the mode switch may experience packet loss if no countermeasures are taken. Thus, it is possible to investigate new architectural solutions to improve the performance of this operation [10]. To enable dynamic mode switching, the following lines must be added to the `omnetpp.ini`, where the `d2dModeSelectionType` parameter contains the name of the module implementing the mode selection algorithm.

```
*.eNodeB.nic.d2dModeSelection = true
*.eNodeB.nic.d2dModeSelectionType="D2DModeSelectionBestCqi"
```

### B. One-to-many D2D

With reference to the scenario of Figure 6, we consider one sender UE, `ueD2D[0]` and two receiving UEs, namely `ueD2D[1]` and `ueD2D[2]`. Multicast messages are generated at the application layer of the node and sent towards a multicast IP address. Thus, they can be received only by UEs that subscribed to the addressed IP multicast group. The latter has to be defined within the XML configuration file of the `IPv4NetworkConfigurator` module. In the example of Figure 6, this is accomplished as follows, where all UEs belong to the group having address 224.0.0.10.

```
<multicast-group hosts="ueD2D[*]"
    interfaces="wlan" address="224.0.0.10"/>
```

In order to let `ueD2D[0]` send multicast messages, the destination IP address of packets generated by its application layer must be set to that value in `omnetpp.ini` file.

```
*.ueD2D[0].udpApp[*].destAddress = "224.0.0.10"
```

The configuration of the `omnetpp.ini` file presents just few changes w.r.t. the one-to-one case described before. However, for one-to-many communications, CQI can be configured only in fixed mode, i.e. setting the `usePreconfiguredTxParams`. The selected CQI affects the transmission range of the communication. In fact, smaller CQIs allow the transmission to be received at farther distances at the cost of higher resource consumption, as more RBs are required to send a message. This can be exploited, for example, in vehicular networks for broadcasting traffic and/or collision information, even in multihop scenarios. In this case, different CQIs allow the packet to reach a different number of vehicles, affecting the performance of the dissemination in terms of latency and resource consumption [11].

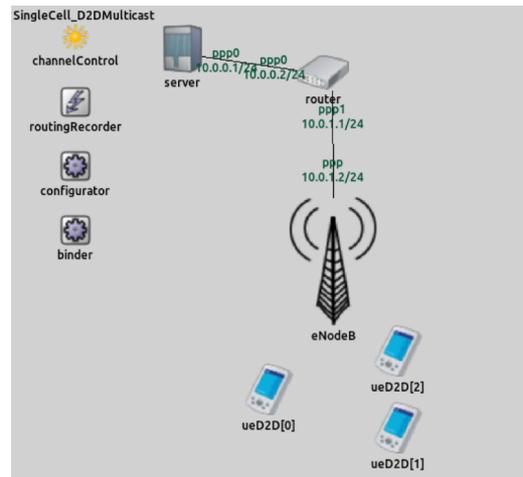

Fig. 6. Example scenario for one-to-many D2D communications

## V. CONCLUSIONS

In this work, we presented SimuLTE as a simulation tool for D2D communications in LTE-Advanced networks. In particular, we provided insight on how to configure simulation parameters to assess the impact of different factors on D2D systems. The presented framework can be exploited, for example, to compare the performance of D2D with other wireless technologies, like Wi-Fi, or to evaluate any kind of network system (e.g. vehicular networks) that might benefit from using one-to-one or one-to-many D2D communications.